\documentclass[12pt]{article}
\usepackage{amsmath,amssymb,bm,graphicx}
\begin{document}

\begin{flushright}
\end{flushright}

\vskip 0.5 truecm

\begin{center}
{\Large{\bf Uncertainty relation and probability: Numerical illustration}}
\end{center}
\vskip .5 truecm
\centerline{\bf  Kazuo Fujikawa$^1$ and Koichiro Umetsu$^{2}$}
\vskip .4 truecm
\centerline {\it $^1$ Institute of Quantum Science, College of 
Science and Technology}
\centerline {\it Nihon University, Chiyoda-ku, Tokyo 101-8308, 
Japan}
\vspace{0.3cm}
\centerline {\it $^2$ Maskawa Institute for Science and Culture, }
\centerline {\it Kyoto Sangyo University, Kita-ku, Kyoto, 603-8555, Japan }
\vskip 0.5 truecm

\makeatletter
\@addtoreset{equation}{section}
\def\theequation{\thesection.\arabic{equation}}
\makeatother

\begin{abstract}
The uncertainty relation and the probability interpretation of quantum mechanics are intrinsically connected, as is evidenced by the evaluation of standard deviations. It is thus natural to ask if one can associate a very small uncertainty product of suitably sampled events with  a very small probability. We have shown elsewhere that 
some examples of the evasion of the uncertainty relation  noted in the past are in fact understood in this way.  We here numerically illustrate that a very small uncertainty product is realized if one performs a suitable sampling of  measured data which occur with a very small probability. It is also shown that our analysis is consistent with the Landau-Pollak type uncertainty relation. It is suggested that the present analysis may help reconcile the contradicting views about the ``standard quantum limit'' in the detection of gravitational waves.

\end{abstract}


\section{Introduction}

The uncertainty relation of Heisenberg~\cite{heisenberg} and an associated detailed analysis of   measurement process \cite{neumann} have been the subjects of main interest for many years. See, for example, references 
\cite{appleby1998} \cite{ozawa2004} \cite{busch2007} \cite{miyadera2008}  
 for the recent analyses of this basic issue. On the other hand, the formulation of the uncertainty relation in the manner of Kennard\cite{kennard} and Robertson\cite{robertson}, which is based only on the commutation relations and the positive metric in the Hilbert space, is straightforward. The uncertainty relation of Kennard evaluates the standard deviations of coordinate and momentum for a given quantum state and thus it is exact, although no direct reference to measurement. In this paper we  study an interrelation between uncertainty and  probability in quantum mechanics by taking the Kennard relation as a basis of the analysis.
 
 Following Heisenberg, it is customary to take the uncertainty relation as a principle, namely, {\em uncertainty principle} which defines the quantum theory at the deepest level. From this point of view, it is impossible to evade the uncertainty relation in the framework of quantum theory. 
However, several authors argued in the past that the evasion of the 
uncertainty relation to an arbitrary degree is possible. For example, Ballentine gave a simple example of the  evasion in the diffraction process\cite{ballentine} and Ozawa gave two simple gedanken experiments\cite{ozawa2003,ozawa2004} which exhibit the evasion of the uncertainty relation. 

The uncertainty relation and the probability interpretation of quantum mechanics are intrinsically intertwined, as is evidenced by the evaluation of the standard deviation. It may thus be natural to incorporate the notion of probability in the study of the uncertainty relation. In fact, we have recently analyzed the basic mechanism involved in the evasion of the uncertainty relation  suggested by the above authors\cite{ballentine,ozawa2003,ozawa2004} from the point of view of probability and uncertainty. We clarified several characteristic features of the evasion of the uncertainty relation, and we have shown that the evasion of the uncertainty relation noted by these authors takes place with a very  small probability\cite{fujikawa2008}.

The {\em sampling} of partial events with preferred properties or a biased measurement of preferred events is important in our analysis. The expectation is that a suitable sampling of the events with preferred properties for an ensemble of similarly prepared states can give a very small uncertainty product $\tilde{\Delta} x \tilde{\Delta} p$, where $\tilde{\Delta} x $ and $\tilde{\Delta} p$ are the standard deviations evaluated for the suitably sampled events, although the 
probability of sampling such events is very small. Classically, this kind of analysis is straightforward. But in quantum mechanics, where  the notion of reduction plays an essential role, this analysis is more involved.    
If one measures the momentum in the preferred range, for example, the quantum state makes a transition to a new state and thus the original information about the coordinate is lost. This aspect is often described as ``measurement creates a quantum state''. 
 
Two aspects of reduction are important in our analysis. In the measurement in quantum mechanics, it is natural to presume an ensemble of similarly prepared states. When one measures the momentum, for example,
each measurement gives a definite value of momentum but the repeated measurement of momentum gives the distribution predicted by quantum mechanics. Similarly the measurement of the coordinate, and the product of the standard deviations of momentum and coordinate thus constructed satisfies the Kennard relation. 
From the point of view of the prepared state, one may be able to assign a definite 
probability to each measured value of the momentum, for example. One may collect only those partial events which occur with  very small probability and form an uncertainty product. The uncertainty product may then turn out to be very small compared to the lower bound of the Kennard relation. From our point of view,  the evasion of the uncertainty relation noted in 
\cite{ballentine} and \cite{ozawa2003,ozawa2004} is an attempt to give a physical meaning to this class of analysis.  

Another aspect of reduction which plays an important role in our analysis is the creation of a new quantum state by measurement. If one measures a specific value of coordinate with high accuracy, the initial state makes a  transition to a new state. One may then imagine an immediately 
subsequent measurement of the specific momentum in the range which is characteristic to the initial state. By this way, one comes back very close to the initial state with a net outcome of the measured values of coordinate and momentum whose uncertainty product is much smaller than 
the lower bound of the Kennard relation.

From the above discussion, it is obvious that we assume the standard interpretation of quantum mechanics. Our attempt is to see if one can find a new aspect in the interplay of uncertainty and probability in quantum mechanics. In the present paper, we present the numerical illustration of the analysis outlined above.   As a possible practical implication of our analysis, it is suggested that 
our analysis may help reconcile the contradicting views on the issue of
standard quantum limit in the detection of gravitational waves~\cite{maddox}.

\section{Uncertainty relation and probability}

We start with a more quantitative analysis of the uncertainty relation and probability. Suppose that we have a suitable localized wave packet $\psi(t,x)$ defined in the one-dimensional space $-\frac{L}{2}\leq x\leq\frac{L}{2}$. We then evaluate the standard deviations of coordinate
$\Delta x$ and momentum $\Delta p$ by using the localized wave packet $\psi(t,x)$. We have the Kennard relation
\begin{eqnarray}
\Delta x\Delta p\geq \frac{1}{2}\hbar
\end{eqnarray}
which is exact.  We take the Kennard relation as a basis of our analysis.
To assign an operational meaning to the Kennard relation, we assume 
a large ensemble of similarly prepared systems.  
We then understand $\Delta x$, for example, as the standard deviation of the coordinate measured by an ideal position detector for an ensemble of states represented by $\psi(t,x)$. Similarly we construct $\Delta p$, and the product of $\Delta x$ and $\Delta p$ thus constructed satisfies the Kennard relation. See, for example, \cite{ballentine}.  

To introduce the notion of probability, we expand the above state as 
\begin{eqnarray}
\psi(t,x)=\sum_{k=1}^{N} c_{k}\phi_{k}(t,x)
\end{eqnarray}
in terms of an orthonormal basis set $\{\phi_{k}(t,x)\}$ where 
each $\phi_{k}(t,x)$ has a support  in $-\frac{1}{2}L+(k-1)(L/N)\leq x\leq -\frac{1}{2}L+k(L/N)$ with $k=1,2,..., N$. By choosing 
$N$ large, one may regard each $\phi_{k}(t,x)$ as an approximate 
eigenstate of the coordinate.
We now repeat the measurement of the standard deviation for the state $\psi(t,x)$ but with $N$ small coordinate detectors (of size $L/N$) placed at the positions of each state  $\phi_{k}(t,x)$. The coordinate detector is triggered only when the particle arrives at the detector. If one collects all the data measured by any of the detectors, one recovers the original value of $\Delta x$. We assign a unit probability to this sampling of the data since we have the same number of measured data as
the number of the similarly prepared states. 

On the other hand, if one collects only the data measured by the specific 
detector corresponding to $\phi_{k_{0}}(t,x)$ one has the standard deviation 
\begin{eqnarray}
\tilde{\Delta} x \sim \frac{1}{N}\Delta x 
\end{eqnarray}
which is evaluated by using the state $\phi_{k_{0}}(t,x)$ for sufficiently large $N$. The quantum mechanical probability for the occurrence of these events is 
\begin{eqnarray}
|c_{k_{0}}|^{2} \sim \frac{1}{N}
\end{eqnarray}
if one normalizes the state $\psi(t,x)$ by $\int_{-L/2}^{L/2}|\psi(t,x)|^{2}dx=\sum_{k}|c_{k}|^{2}=1$. We thus {\em assign the notion of probability to each data set}. From this definition, one sees that our probability is a relative probability rather than the absolute probability.
We can consider the similar construction for the momentum measurement of the state $\psi(t,x)$. 

If one considers the case where all the momentum measurements are accepted
but only the coordinate measured by the specific detector corresponding to $\phi_{k_{0}}(t,x)$ is accepted for the prepared state 
 $\psi(t,x)$, one has
an analogue of the Kennard relation
\begin{eqnarray}
\Delta p\tilde{\Delta} x \sim \frac{1}{N}\Delta x\Delta p \sim \frac{1}{N}\hbar\ll \hbar.
\end{eqnarray}
The quantum mechanical probability for this sampling of events for the ensemble of states represented by $\psi(t,x)$ is given by (2.4), which is very small.

It is important to realize that the above uncertainty product (2.5)
is also the natural product when one measures only the coordinate by the above specific detector but no measurement of the momentum for the given initial state $\psi(t,x)$. This is the typical situation of the {\em partial {\rm (} i.e., only the coordinate or momentum is directly measured{\rm )} or indirect {\rm (} i.e., either the momentum or coordinate distribution is theoretically guessed{\rm )} measurement}. If one knows the prepared initial state, one may guess the uncertainty in the momentum as the standard deviation as in (2.5) without a direct measurement of the momentum.

 It is shown in Appendix that a small detector limit in the analysis of the evasion of the uncertainty
relation in the diffraction process discussed by Ballentine
\cite{ballentine}, which is based on a partial measurement, precisely corresponds to (2.4) and (2.5).
It is also shown in Appendix that one of the gedanken experiments of Ozawa (see Section 
9 in \cite{ozawa2003}), which evades the uncertainty relation
in the form $\eta(p)\epsilon(x)\ll \hbar$ with the measurement error $\epsilon(x)$
and the disturbance $\eta(p)$, is described by the expansion (2.3) and the probability (2.4). The gedanken experiment of Ozawa is also based on a partial measurement. These facts may suggest that one might call the relations (2.4) and (2.5) as ``an evasion of the uncertainty relation to an arbitrary degree with very small probability'', although we operate in the framework of standard quantum mechanics and thus do not evade the standard Kennard relation.
   
As for the interpretation of (2.5) as a result of the partial measurement of the prepared state $\psi(t,x)$ by a specific coordinate detector, one may notice that once the state is reduced to    
$\phi_{k_{0}}(t,x)$ the standard deviations of coordinate and 
momentum evaluated for $\phi_{k_{0}}(t,x)$ precisely satisfy the ordinary
Kennard relation. One may thus ask what is the use of the  relation (2.5)?  As an answer to this question, we propose a specific subsequent measurement of the momentum  by expanding $\phi_{k_{0}}(t,x)$ in the form
\begin{eqnarray}
\phi_{k_{0}}(t,x) = \sum_{l}a_{k_{0},l}\varphi_{l}(t,x)
\end{eqnarray}
where an orthonormal set $\{\varphi_{l}(t,x)\}$ consists of localized wave packets  (approximate momentum eigenstates) in the {\em original} interval 
$-L/2\leq x\leq L/2$.

Our next gedanken experiment is to collect only the data corresponding to  the momentum belonging to a specific state $\varphi_{l_{0}}(t,x)$ in (2.6) in the measurement of the reduced state $\phi_{k_{0}}(t,x)$, which is performed immediately after the measurement of initial $\psi(t,x)$ by the above specific coordinate detector. The above specific coordinate measurement may now be regarded as a preparation of the state $\phi_{k_{0}}(t,x)$, and thus the present momentum measurement is also a partial measurement. We choose the state $\varphi_{l_{0}}(t,x)$ which is close to the 
starting state $\psi(t,x)$; it is shown later that this is possible by choosing the starting state $\psi(t,x)$ suitably.
In this sampling of the data of the momentum measurement, the standard  deviation of the momentum $\tilde{\Delta}p$, which is actually evaluated by using the state $\varphi_{l_{0}}(t,x)$, is given by 
\begin{eqnarray}
\tilde{\Delta}p\sim \Delta p
\end{eqnarray}
where $\Delta p$ is the standard deviation for the state $\psi(t,x)$ in (2.1). The uncertainty product of the standard deviation of coordinate in the preparation process of $\phi_{k_{0}}(t,x)$ and the standard deviation of momentum in the immediately subsequent 
measurement of the momentum corresponding to the state  $\varphi_{l_{0}}(t,x)$ is then given by 
\begin{eqnarray}
\tilde{\Delta}x\tilde{\Delta}p\sim \frac{1}{N}\Delta x\Delta p
\sim \frac{\hbar}{N}\ll \hbar.
\end{eqnarray}
The above specific measurement (or sampling) of momentum creates the state $\varphi_{l_{0}}(t,x)$, and the probability of finding $\varphi_{l_{0}}(t,x)$ in the state  $\phi_{k_{0}}(t,x)$  in (2.6) is 
\begin{eqnarray}
|a_{k_{0},l_{0}}|^{2}\sim \frac{1}{N}.
\end{eqnarray}
The net outcome of this approximate ``cyclic measurements''
$\psi(t,x)\rightarrow \phi_{k_{0}}(t,x)\rightarrow \varphi_{l_{0}}(t,x)$ with $\varphi_{l_{0}}(t,x)\sim \psi(t,x)$ is the relation (2.8), although
such a probability is very small; the intrinsic quantum probability for the occurrence of (2.8) is $\sim 1/N$ as is seen in (2.9), but if one recalls that one started 
with an ensemble of states represented by $\psi(t,x)$, the probability 
to arrive at the final state $\varphi_{l_{0}}(t,x)$ by two steps is 
$\sim 1/N^{2}$. 

Obviously, this  ``cyclic measurements'' differs from the ``simultaneous measurements'' of  coordinate and momentum for the 
state $\psi(t,x)$, but one can extract the information about coordinate and momentum which gives a very small uncertainty product in (2.8) by restoring the state $\psi(t,x)$ approximately to its original form.

Our suggestion is that the relations (2.8) and (2.9) might have some bearing on the analysis of the ``standard quantum limit'' in the 
detection of gravitational waves\cite{maddox}. The basic issue   
in the detection of gravitational waves is the accurate measurement of the coordinate and then how to control the subsequent time development of the system. This time development of the system is controlled by the fluctuation of the momentum after the coordinate measurement. We make a further comment on this issue in Section 5. 
\\
\\
{\bf Comparison with the Landau-Pollak type uncertainty relation}
\\

We here show that our analysis is consistent with the Landau-Pollak type uncertainty relation which states that 
\begin{eqnarray}
\langle\eta|E|\eta\rangle+\langle\eta|P|\eta\rangle\leq 1+ ||EP|| 
\end{eqnarray}
for two projection operators $E$ and $P$, and any normalized state $|\eta\rangle$~\cite{miyadera2007}. The Landau-Pollak type relation also emphasizes the probability aspect of the uncertainty relation.
If one chooses
\begin{eqnarray}
E=\int_{x_{0}-\frac{1}{2}\delta x}^{x_{0}+\frac{1}{2}\delta x}dx |x\rangle\langle x|, \ \  
P=\int_{p_{0}-\frac{1}{2}\delta p}^{p_{0}+\frac{1}{2}\delta p}\frac{dp}{2\pi\hbar} |p\rangle\langle p|, 
\end{eqnarray}
one has
\begin{eqnarray}
||EP||^{2}=||EPE||\leq {\rm Tr}(EPE)={\rm Tr}(PEP)
\end{eqnarray}
and 
\begin{eqnarray}
{\rm Tr}(EPE)&=&\int_{x_{0}-\frac{1}{2}\delta x}^{x_{0}+\frac{1}{2}\delta x}dx\int_{p_{0}-\frac{1}{2}\delta p}^{p_{0}+\frac{1}{2}\delta p}\frac{dp}{2\pi\hbar}\int_{x_{0}-\frac{1}{2}\delta x}^{x_{0}+\frac{1}{2}\delta x}dx^{\prime}  \langle x^{\prime}|x\rangle\langle x|p\rangle\langle p|x^{\prime}\rangle\nonumber\\
&=& \int_{x_{0}-\frac{1}{2}\delta x}^{x_{0}+\frac{1}{2}\delta x}dx\int_{p_{0}-\frac{1}{2}\delta p}^{p_{0}+\frac{1}{2}\delta p}\frac{dp}{2\pi\hbar} \langle x|p\rangle\langle p|x\rangle\nonumber\\
&=&\int_{x_{0}-\frac{1}{2}\delta x}^{x_{0}+\frac{1}{2}\delta x}dx\int_{p_{0}-\frac{1}{2}\delta p}^{p_{0}+\frac{1}{2}\delta p}\frac{dp}{2\pi\hbar} e^{ipx/\hbar}e^{-ipx/\hbar}\nonumber\\
&=&\frac{\delta x\delta p}{2\pi\hbar}.
\end{eqnarray}
The inequality (2.10) implies that either  $\langle\eta|E|\eta\rangle$ or $\langle\eta|P|\eta\rangle$ (or both) is forced to be significantly smaller than unity when
\begin{eqnarray}
\frac{\delta x\delta p}{2\pi\hbar}\ll 1.
\end{eqnarray}

From this point of view, the relations (2.4) and (2.5) in our analysis are regarded to correspond to the choice of  a specific wave packet $\psi(x)=\langle x|\psi\rangle=\int_{p_{0}-\frac{1}{2}\Delta p}^{p_{0}+\frac{1}{2}\Delta p}\frac{dp}{2\pi\hbar}\langle x|p\rangle\langle p|\psi\rangle=\int_{p_{0}-\frac{1}{2}\Delta p}^{p_{0}+\frac{1}{2}\Delta p}\frac{dp}{2\pi\hbar}e^{ipx/\hbar}\langle p|\psi\rangle$ with $P|\psi\rangle=|\psi\rangle$ and $\langle\psi|P|\psi\rangle=\langle\psi|\psi\rangle=1$, namely, $P=\int_{p_{0}-\frac{1}{2}\Delta p}^{p_{0}+\frac{1}{2}\Delta p}\frac{dp}{2\pi\hbar}|p\rangle\langle p|$. Then 
\begin{eqnarray}
\langle\psi|E|\psi\rangle=\langle\psi|PEP|\psi\rangle
\leq ||PEP||\leq {\rm Tr}(PEP)\simeq\frac{\tilde{\Delta}x\Delta p}{2\pi\hbar} 
\end{eqnarray}
and the small probability $\langle\psi|E|\psi\rangle$ with $E=\int_{x_{0}-\frac{1}{2}\tilde{\Delta}x}^{x_{0}+\frac{1}{2}\tilde{\Delta}x}dx |x\rangle\langle x|$
corresponds to (2.4).
This inequality (2.15) is more stringent than the weak version of the Landau-Pollak type uncertainty relation (2.10) formulated by Miyadera and Imai~\cite{miyadera2007}, which contains the square root of (2.14) as the 
upper bound.

Similarly, the relations (2.8) and (2.9) are regarded to correspond to the choice of a specific state 
$\langle\phi|E|\phi\rangle=\langle\phi|\phi\rangle=1$ with $E=\int_{x_{0}-\frac{1}{2}\tilde{\Delta}x}^{x_{0}+\frac{1}{2}\tilde{\Delta}x}dx |x\rangle\langle x|$
, and 
\begin{eqnarray}
\langle\phi|P|\phi\rangle=\langle\phi|EPE|\phi\rangle
\leq ||EPE||\leq {\rm Tr}(EPE)\simeq\frac{\tilde{\Delta}x\tilde{\Delta}p}{2\pi\hbar}
\end{eqnarray}
for $P=\int_{p_{0}-\frac{1}{2}\tilde{\Delta} p}^{p_{0}+\frac{1}{2}\tilde{\Delta} p}\frac{dp}{2\pi\hbar}|p\rangle\langle p|$.
The left-hand side $\langle\phi|P|\phi\rangle$ of this inequality gives the small probability corresponding to our relation (2.9) when the upper bound in (2.16) is small. See also (3.31) and (3.32) in Section 3. 

 In reality, the actual spreads of coordinate and momentum in the projection operators in (2.11) are larger than the standard deviations to satisfy the condition such as $P|\psi\rangle=|\psi\rangle$ for a given $|\psi\rangle$, and thus the precise upper bound is expected to be larger than the values in (2.15)
 and (2.16) by some finite factor.
These inequalities (2.15) and (2.16) are useful when the upper bound is significantly smaller than unity.

\section{ Procedure of the numerical calculation}

In this section, we describe the procedure of the numerical calculation,
and the detailed numerical evaluation itself is presented in Section 4.

For the numerical illustration of a very small uncertainty product for a suitably sampled date set, we consider the simplest Schr\"{o}dinger equation
\begin{eqnarray}
i\hbar\frac{\partial}{\partial t}\psi(x,t)=H\psi(x,t)
\end{eqnarray}
with
\begin{eqnarray}
H=\frac{\hat{p}^{2}}{2m}=-\frac{\hbar^{2}}{2m}\frac{\partial^{2}}{\partial x^{2}}
\end{eqnarray}
in a one-dimensional box with a size $L$ ($0\leq x\leq L$) and with the periodic boundary condition. Then the basic solution is
\begin{eqnarray}
\psi_{n}(x,t)=\frac{1}{\sqrt{L}}\exp{[i\frac{p_{n}x}{\hbar}-i\frac{p_{n}^{2}}{2m\hbar}t]}
\end{eqnarray}
with $p_{n}=2\pi\hbar n/L, \ \ \ n=0, \pm 1, \pm 2,....\ $, but the pure plane wave which is an eigenstate of momentum causes complications in the analysis of the standard form of the Kennard relation since it leads to $\Delta p \Delta x\geq 0$. To ensure the standard Kennard relation, we exploit the fact that any free particle created in the laboratory is localized in space. The readers are asked to refer to Ref.~\cite{fujikawa2010} for the technical details of the procedure used in the present paper. We thus consider the wave packets, which are actually the superposition of two plane wave solutions,~\cite{fujikawa2010} 
\begin{eqnarray}
\psi_{n,k}(x,t)&\equiv&\exp[i\frac{p_{n}x}{\hbar}-i\frac{p_{n}^{2}}{2m\hbar} t]\nonumber\\
&&\hspace{1cm}\times\frac{2i}{\sqrt{2L}}\exp[-i\frac{p_{k}^{2}}{2m\hbar} t]
\sin\left(\frac{p_{k}}{\hbar}[x-(p_{n}/m) t]\right)
\end{eqnarray}
defined in the interval $0\leq x\leq L$ at $t=0$ where
\begin{eqnarray}
p_{n}\equiv\frac{\pi\hbar }{L}n
\end{eqnarray}
with integer $n$, and similarly $p_{k}$.  This construction, when looked at $t=0$, is analogous to the Bloch wave with the Bloch momentum $p_{n}$ and a complete set of sine functions in the interval $0\leq x\leq L$ if one considers all positive integers $k$: To be precise, we have periodic wave packets in the extended interval $-L\leq x\leq L$ due to the presence of periodic (even $k$) and anti-periodic (odd $k$) waves in the interval $0\leq x\leq L$, but we use only half of them defined in $0\leq x\leq L$ at $t=0$~\footnote{This is consistent since the probability flow at the points $x=(p_{n}/m) t$ and $x=L+(p_{n}/m) t$ up to a multiple of $2L$ is always zero for any time $t$.}. This construction, which is analogous to the Bloch wave, allows us to introduce the zero in the wave function (i.e., locality) to ensure the ordinary Kennard relation and at the same time to retain the notion of  momentum related to the plane wave~\cite{fujikawa2010}.

Both of the solutions (3.2) and (3.4) satisfy the Schr\"{o}dinger equation (3.1), but the difference is that the wave packet in (3.4) is actually moving with the velocity $p_{n}/m$. Any free particle with an initial momentum $\langle p\rangle=\tilde{p}_{n}$ localized in the sub-domain of $[0,L]$ at $t=0$ is expanded as  
\begin{eqnarray}
\psi(x,t)=\sum_{k=1}c_{k}\psi_{\tilde{n},k}(x,t)
\end{eqnarray}
where $\psi_{\tilde{n},k}(x,t)$ stands for the wave packet in (3.4) with
$p_{n}$ replaced by $\tilde{p}_{n}$, and for general $\tilde{p}_{n}$
\begin{eqnarray}
\psi(x+2L,t)=\exp[i\frac{2\tilde{p}_{n}L}{\hbar}]\psi(x,t)
\end{eqnarray}
which is the Bloch-like periodicity condition. The solution (3.6) is written as 
\begin{eqnarray}
\psi(x,t)=\exp[i\frac{\tilde{p}_{n}x}{\hbar}-i\frac{\tilde{p}_{n}^{2}}{2m\hbar} t]\phi(x-(\tilde{p}_{n}/m) t, t)
\end{eqnarray}
where $\phi(x,t)$ formally corresponds to a solution of a free particle confined 
in a deep potential well with a width $0\leq x\leq L$,
\begin{eqnarray}
\phi(x,t)&=&\sum_{k=1}c_{k}\frac{2i}{\sqrt{2L}}\exp[-i\frac{p_{k}^{2}}{2m\hbar} t]\sin\left(\frac{k\pi}{L}x\right).
\end{eqnarray}
In our case, however, the deep potential well is moving with the velocity $(\tilde{p}_{n}/m)$.
Not only each wave in (3.4) but also any superposition of the waves such as in (3.6) satisfy the condition on the circle $S^{1}$ with a circumference $2L$ (which is a natural domain for the periodic boundary condition)
\begin{eqnarray}
{\rm Min}_{x\in S^{1}}|\psi(x,t)|^{2}=0
\end{eqnarray}
for any $t$, namely at $x=(\tilde{p}_{n}/m) t$ and $x=L+(\tilde{p}_{n}/m) t$ up to a multiple of $2L$. The condition (3.10) is the {\em locality requirement} in our formulation. One can thus define the Kennard relation~\cite{fujikawa2010}
\begin{eqnarray}
\Delta p \Delta x\geq \frac{\hbar}{2}
\end{eqnarray}
where the integration domain to evaluate $\Delta p$ and $\Delta x$ is taken to be $[(\tilde{p}_{n}/m) t, L + (\tilde{p}_{n}/m) t]$. 
This construction may appear to be a technical detail, but it is essential for a reliable analysis of the magnitude of the uncertainty product in connection with the Kennard relation.  

For the elementary solution $\psi_{n,k}(x,t)$ in (3.4) we have
\begin{eqnarray}
\langle p\rangle&=&\int_{(p_{n}/m) t}^{L+(p_{n}/m) t}\psi^{\star}_{n,k}(x,t)\frac{\hbar}{i}\partial_{x}\psi_{n,k}(x,t)dx\nonumber\\
&=&p_{n}\nonumber\\
\langle p^{2}\rangle&=&\int_{(p_{n}/m) t}^{L+(p_{n}/m) t}\psi^{\star}_{n,k}(x,t)(\frac{\hbar}{i}\partial_{x})^{2}\psi_{n,k}(x,t)dx\nonumber\\
&=&\frac{1}{2}[(p_{n}+p_{k})^{2}+(p_{n}-p_{k})^{2}]\nonumber\\
\Delta p&=&\sqrt{p_{k}^{2}}=k\frac{\pi\hbar}{L}
\end{eqnarray}
and
\begin{eqnarray}
\langle x\rangle&=&\frac{2}{L}\int_{(p_{n}/m) t}^{L+(p_{n}/m) t}x\sin^{2}\left(k\frac{\pi}{L}[x-(p_{n}/m) t]\right)dx\nonumber\\
&=&\frac{L}{2}+\frac{p_{n}}{m} t,\nonumber\\
\langle x^{2}\rangle&=&\frac{2}{L}\int_{(p_{n}/m) t}^{L+(p_{n}/m) t}x^{2}\sin^{2}\left(k\frac{\pi}{L}[x-(p_{n}/m) t]\right)dx\nonumber\\
&=&\frac{L^{2}}{3}-\frac{2L^{2}}{(2\pi k)^{2}}+2(\frac{L}{2})(\frac{p_{n}}{m} t) +(\frac{p_{n}}{m} t)^{2}, \nonumber\\
\Delta x&=&\frac{L}{2\sqrt{3}}\sqrt{1-\frac{24}{(2\pi k)^{2}}}.
\end{eqnarray}
Thus
\begin{eqnarray}
\Delta x\Delta p=\frac{\pi\hbar}{\sqrt{3}}\sqrt{k^{2}-\frac{24}{(2\pi )^{2}}}.
\end{eqnarray}
The choice $k=1$ gives the minimum uncertainty state in our construction, and we have
\begin{eqnarray}
\Delta x\Delta p=\frac{\pi\hbar}{2\sqrt{3}}\sqrt{1-\frac{24}{(2\pi )^{2}}}\ >\frac{\hbar}{2}.
\end{eqnarray}
The numerical value of the uncertainty product 
$\Delta x\Delta p$ in (3.15) is close to the lower bound $\frac{\hbar}{2}$.

We thus choose our initial state to analyze the uncertainty relation as 
\begin{eqnarray}
\psi_{n,1}(x,t)
&=&\frac{2i}{\sqrt{2L}}\exp[i\frac{p_{n}x}{\hbar}-i\frac{p_{n}^{2}+p_{1}^{2}}{2m\hbar} t]\nonumber\\
&&\hspace{1cm}\times\sin\left(\frac{\pi}{L}[x-(p_{n}/m) t]\right).
\end{eqnarray}
To perform a numerical analysis, we define dimensionless quantities:
\begin{eqnarray}
&&\bar{x}=\frac{x}{L},\hspace{5mm} 0\leq \bar{x}\leq 1   \nonumber\\
&&\bar{p}_{n}=\frac{\pi\hbar n}{L}/(\frac{\hbar}{L})=\pi n,\nonumber\\
&&\lambda=\lambda_{c}/L=(\frac{\hbar}{mc})/L,\nonumber\\
&&T=ct/L.
\end{eqnarray}
Then the above wave packet (3.16) is written as 
\begin{eqnarray}
\psi_{n,1}(\bar{x},T)
&=&\frac{2i}{\sqrt{2}}\exp[i\bar{p}_{n}\bar{x}-i\frac{\bar{p}_{n}^{2}+\bar{p}_{1}^{2}}{2}\lambda T]\nonumber\\
&&\hspace{1cm}\times\sin\left(\pi[\bar{x}-\bar{p}_{n}\lambda T]\right).
\end{eqnarray}
and the standard Kennard relation is given by 
\begin{eqnarray}
\Delta\bar{x}\Delta\bar{p} \geq \frac{1}{2}.
\end{eqnarray}

We now describe 4 steps in our numerical analysis of the measurement process:  For a notational simplicity, we choose the time of our measurements at $T=0$.\\

\noindent (i) We start with the normalized wave packet (3.18) which satisfies the 
standard Kennard relation (3.19). We accept all the measured events in the evaluation of the standard deviations in (3.19).
This sampling of events takes place with a {\em unit probability} (by our definition of probability) for an ensemble of similarly prepared states represented by the wave packet $\psi_{n,1}(\bar{x},0)$.\\
\\
(ii) Next suppose to sample only those events measured by the specific position detector, of which size $a$ is much
smaller than the size of the wave packet (and also the size of the box) $L$, for an ensemble of similarly prepared states represented by the above wave packet $\psi_{n,1}(\bar{x},0)$. Namely, 
\begin{eqnarray}
N\equiv\frac{L}{a}\gg 1
\end{eqnarray}
and we choose $N$ to be an integer. 

We introduce a set of normalized step functions $\{u_{l}(\bar{x})\}$
by 
\begin{eqnarray}
u_{l}(\bar{x})=\sqrt{N}, \hspace{1cm} (l-1)/N\leq \bar{x} \leq l/N
\end{eqnarray}
and $u_{l}(\bar{x})=0$ otherwise, for $l=1,2,...,N$. 
One may then recognize that the original wave packet in (3.18) is written as 
\begin{eqnarray}
\psi_{n,1}(\bar{x},0)=[ \frac{1}{\sqrt{N}}\sum_{l=1}^{N}u_{l}(\bar{x}) ]\psi_{n,1}(\bar{x},0)
=\sum_{l=1}^{N}c_{l}\phi_{l}(\bar{x},0)
\end{eqnarray}
where   
\begin{eqnarray}
&&\phi_{l}(\bar{x},0)\equiv
\psi_{n,1}(\bar{x},0)u_{l}(\bar{x})/\sqrt{B_{l}},\nonumber\\
&&c_{l}=\sqrt{\frac{B_{l}}{N}},
\end{eqnarray}
with 
\begin{eqnarray}
B_{l}=\int_{0}^{1}d\bar{x}|\psi_{n,1}(\bar{x},0)u_{l}(\bar{x})|^{2}.
\end{eqnarray}
The set $\{\phi_{l}(\bar{x},0)\}$ forms an {\em orthonormal set} in the 
interval $0 \leq \bar{x}\leq 1$, and each $\phi_{l}(\bar{x},0)$ has a support in $(l-1)/N\leq \bar{x} \leq l/N$. 

We now regard that the measurement of coordinate by the small position detector described above corresponds to picking up a specific state $\phi_{l_{0}}(\bar{x},0)$. This means that we make a very specific sampling of events corresponding to the state $\phi_{l_{0}}(\bar{x},0)$ for the prepared state $\psi_{n,1}(\bar{x},0)$. The standard deviation of coordinate in this sampling is given by 
\begin{eqnarray}
\tilde{\Delta} \bar{x}\sim \Delta \bar{x}_{l_{0}}   \ll \Delta\bar{x}
\end{eqnarray}
where $\Delta \bar{x}_{l_{0}}$ is evaluated by using the state $\phi_{l_{0}}(\bar{x},0)$. We assign the probability 
\begin{eqnarray}
|c_{l_{0}}|^{2}=\frac{B_{l_{0}}}{N} \ll 1
\end{eqnarray}
to this specific sampling of measured coordinate, which corresponds to the 
reduction probability of the state $\psi_{n,1}(\bar{x},0)$ to  $\phi_{l_{0}}(\bar{x},0)$ in the expansion (3.22). This probability is also written as 
\begin{eqnarray}
|c_{l_{0}}|^{2}=\int_{(l_{0}-1)/N}^{l_{0}/N}d\bar{x}|\psi_{n,1}(\bar{x},0)|^{2}. 
\end{eqnarray}
The state 
$\phi_{l_{0}}(\bar{x},0)$ after the specific measurement generally depends on the initial state $\psi_{n,1}(\bar{x},0)$, but this dependence diminishes when one chooses $N\gg 1$.

If one assumes that all the events in the momentum measurement of the original wave packet $\psi_{n,1}(\bar{x},0)$ are accepted, then the standard deviation of the momentum is given by $\Delta \bar{p}$ in (3.19).
The uncertainty product for this specific sampling of events then becomes  
\begin{eqnarray}
\Delta \bar{p}\tilde{\Delta} \bar{x}\sim \Delta \bar{p}\Delta \bar{x}_{l_{0}} \ll \frac{1}{2}
\end{eqnarray}
while the probability for this sampling of events is given by (3.26). 
\\
\\
(iii) We next suppose to collect all the measured data of momentum for the above {\em reduced} state $\phi_{l_{0}}(\bar{x},0)$
without any restriction on the value of momentum. It is then confirmed that the Kennard relation holds
\begin{eqnarray}
\Delta \bar{p}_{l_{0}}\Delta \bar{x}_{l_{0}} \geq \frac{1}{2}
\end{eqnarray}
where $\Delta \bar{p}_{l_{0}}$ is calculated by using the localized state $\phi_{l_{0}}(\bar{x},0)$.
We assign a {\em unit probability} to this sampling of data for the prepared states represented by $\phi_{l_{0}}(\bar{x},0)$. The relation (3.29) is what one naively expects; the precise measurement of coordinate leads to the spread momentum.\\
\\
(iv) For the reduced wave function $\phi_{l_{0}}(\bar{x},0)$, we next suppose to {\em selectively measure} the specific momentum, namely, we sample only the events with momentum  which {\em approximately} corresponds to the original wave packet $\psi_{n,1}(\bar{x},0)$ in (3.18) in the expansion (see the expansion in (3.6))
\begin{eqnarray}
&&\phi_{l_{0}}(\bar{x},0)=\sum_{k=1}a_{l_{0}, k}\psi_{n,k}(\bar{x},0),\nonumber\\
&&a_{l_{0}, k}=\int_{0}^{1}\psi^{\dagger}_{n,k}(\bar{x},0)\phi_{l_{0}}(\bar{x},0)d\bar{x}.
\end{eqnarray}
In this procedure one may regard the specific coordinate measurement in the analysis (ii) as a preparation of the state $\phi_{l_{0}}(\bar{x},0)$ with the standard deviation $\Delta \bar{x}_{l_{0}}$, and the present immediately subsequent measurement as an analysis of the state $\phi_{l_{0}}(\bar{x},T)$ by a specific momentum analyzer. The expected standard deviation of momentum in this specific measurement (or sampling) is $\Delta \bar{p}$, which is the standard deviation for the original wave packet in (3.18), and the uncertainty product is
\begin{eqnarray}
\Delta \bar{p}\Delta \bar{x}_{l_{0}}\ll \frac{1}{2}
\end{eqnarray}
which is  identical to the uncertainty product in the case (ii) above. By taking (3.30) into account, we assign a probability \begin{eqnarray}
|a_{l_{0}, 1}|^{2}=|\int_{0}^{1}\psi^{\dagger}_{n,1}(\bar{x},0)\phi_{l_{0}}(\bar{x},0)d\bar{x}|^{2}
&=& |c_{l_{0}}|^{2}
\end{eqnarray}
to the specific sampling of events in (3.31), when it is assumed to be feasible at least approximately.
This probability  agrees with the probability in (3.26).

By this specific measurement (or sampling) of the momentum, we come back close to the original wave function $\psi_{n,1}(\bar{x},0)$ in (3.18).  The net outcome in this approximate cycle is the measurements of the coordinate in the analysis (ii) and the momentum in the analysis (iv) which give the uncertainty product much smaller than the lower bound of the Kennard relation as in (3.31).
 The importance of the analysis (iv) is to show that the analysis (iii), which is the commonly expected result of the precise measurement of the coordinate, is not the end of the story. The approximate restoration to the original state $\psi_{n,1}(\bar{x},0)$ is an application of the {\em creation of a quantum state by measurement}.

\section{Actual numerical calculation}

In this section, we explain some details of the numerical calculation.
We fix the parameter $T=0$ in our analysis for simplicity, and thus the parameter $\lambda$ in (3.17) does not appear in our analysis. For an analysis at 
 $T\neq 0$, one may choose, for example,
$\lambda=10^{-5}$
which means that the size of the box is $10^{5}$ times the Compton wave length of the particle involved.

We choose the starting wave function $\psi_{n,1}(\bar{x},0)$ in (3.18) with $n=10$. To achieve a very small uncertainty product, we choose the detector parameter defined in (3.20) at
$N=100$ or $N=200$, and the position of the detector slightly away from the center of the box, namely, $l_{0}=40$ or $l_{0}=80$, respectively.

For those parameters, we repeat the analyses (i) to (iv) in Section 3. For each case, we checked  $|\phi_{l_{0}}(\bar{x},0)|^{2}$
and the (momentum space) distribution $|a_{l_{0},k}|^{2}$ of $\phi_{l_{0}}(\bar{x},0)$. The distribution $|\psi_{n,1}(\bar{x},0)|^{2}$ for the case $n=10$, which is actually independent of $n$, is shown in Fig.1.
\\
\begin{figure}[ht]
  \begin{center}
    \includegraphics[width=7cm]{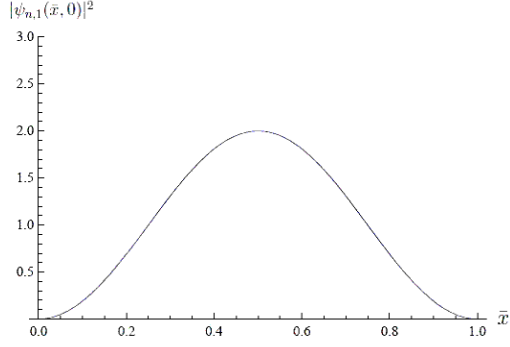}
  \caption{The distribution $|\psi_{n,1}(\bar{x},0)|^{2}$, $0 \leq \bar{x}\leq 1$, which is independent of $n$.}
  \end{center}
\end{figure}

For illustration, we show the details of the numerical calculation for the case $n=10$ and $N=200$ later. The expansion (3.30) is used to evaluate the standard distribution of the momentum for $\phi_{l_{0}}(\bar{x},0)$
\begin{eqnarray}
\langle \bar{p}\rangle_{l_{0}}&=&\int_{0}^{1}d\bar{x}\phi^{\dagger}_{l_{0}}(\bar{x},0)\frac{1}{i}\frac{\partial}{\partial \bar{x}}\phi_{l_{0}}(\bar{x},0)\nonumber\\
&=&\bar{p}_{n}+\int_{0}^{1}\varphi^{\dagger}(\bar{x},0)\frac{1}{i}\frac{\partial}{\partial \bar{x}}\varphi(\bar{x},0)d\bar{x}\nonumber\\
&=&\bar{p}_{n}+\langle \bar{p}\rangle_{\varphi}\nonumber\\
\langle \bar{p}^{2}\rangle_{l_{0}}&=&\int_{0}^{1}d\bar{x}\phi^{\dagger}_{l_{0}}(\bar{x},0)(\frac{1}{i}\frac{\partial}{\partial \bar{x}})^{2}\phi_{l_{0}}(\bar{x},0)\nonumber\\
&=&\bar{p}_{n}^{2}+2\bar{p}_{n}\langle \bar{p}\rangle_{\varphi}+\int_{0}^{1}\varphi^{\dagger}(\bar{x},0)(\frac{1}{i}\frac{\partial}{\partial \bar{x}})^{2}\varphi(\bar{x},0)d\bar{x}\nonumber\\
&=&\bar{p}_{n}^{2}+2\bar{p}_{n}\langle \bar{p}\rangle_{\varphi}+\langle \bar{p}^{2}\rangle_{\varphi}\nonumber\\
(\Delta\bar{p}_{l_{0}})^{2}&=&\langle\bar{p}^{2}\rangle_{\varphi}-\left(\langle \bar{p}\rangle_{\varphi}\right)^{2}
\end{eqnarray}
where we defined
\begin{eqnarray}
\phi_{l_{0}}(\bar{x},0)&=&\sum_{k=1}a_{l_{0}, k}\psi_{n,k}(\bar{x},0)\nonumber\\
&=&e^{i\bar{p}_{n}\bar{x}}\varphi(\bar{x},0),\nonumber\\
\varphi(\bar{x},0)
&\equiv&\sum_{k=1}a_{l_{0}, k}\frac{2i}{\sqrt{2}}
\sin\left(k\pi\bar{x}\right),\nonumber\\
\langle \bar{p}\rangle_{\varphi}&=&\frac{1}{2i}[\int_{0}^{1}\varphi^{\dagger}(\bar{x},0)\frac{\partial}{\partial \bar{x}}\varphi(\bar{x},0)d\bar{x}-h.c.],\nonumber\\
\langle \bar{p}^{2}\rangle_{\varphi}&=&\frac{-1}{2}[\int_{0}^{1}\varphi^{\dagger}(\bar{x},0)(\frac{\partial}{\partial \bar{x}})^{2}\varphi(\bar{x},0)d\bar{x}+h.c.].
\end{eqnarray}
This procedure is convenient to ensure the hermiticity of the momentum 
operator $\hat{\bar{p}}$. However, due to the $\delta$-functional singularity in the derivative of the step-function, the coefficient
$a_{l_{0}, k}$ contains arbitrary large frequency $k$ and it causes the divergence in the above summation such as  $\langle \bar{p}^{2}\rangle_{l_{0}}$ in (4.1). To remedy this divergence introduced by the (artificial) sharp step function $\phi_{l_{0}}(\bar{x},0)$, we cut off the summation in (4.2) at $k= 4N$ in the momentum space, which means a smoothing of the spatial function $\phi_{l_{0}}(\bar{x},0)$. In the actual calculation, we first plot the distribution $|a_{l_{0}, k}|^{2}$ and confirm that this cut-off in $k$ is reasonable.

From the value of the {\em uncertainty product}
\begin{eqnarray}
U=\Delta\bar{p}\Delta\bar{x}
\end{eqnarray}
together with the quantum mechanical (relative) {\em probability} $P$ for each case, one can confirm our statements in Section 3. Note that the standard deviations $\Delta\bar{p}$ and $\Delta\bar{x}$ in (4.3) are generally defined for a specific sampling of measured events and thus generally differ from those appearing in the standard Kennard relation as is explained in Sections 2 and 3.   

The results are:\\
(1) Wave function with the parameter $n=10$ in (3.18) and the detector with the parameter $N=100$ in (3.20):
\begin{eqnarray}
U_{({\rm i})}&=&0.567862, \hspace{0.5cm} P_{({\rm i})}=1,\nonumber\\
U_{({\rm ii})}&=&0.00906856, \hspace{0.5cm} P_{({\rm ii})}=0.0179003,\nonumber\\
U_{({\rm iii})}&=&0.826994, \hspace{0.5cm} P_{({\rm iii})}=1,\nonumber\\
U_{({\rm iv})}&=&0.00906856, \hspace{0.5cm} P_{({\rm iv})}\simeq 0.0179003
\end{eqnarray}
\\
(2) Wave function with the parameter $n=10$ in (3.18) and the detector with the parameter  with $N=200$ in (3.20):
\begin{eqnarray}
U_{({\rm i})}&=&0.567862, \hspace{0.5cm} P_{({\rm i})}=1,\nonumber\\
U_{({\rm ii})}&=&0.00453444, \hspace{0.5cm} P_{({\rm ii})}=0.00899826,\nonumber\\
U_{({\rm iii})}&=&0.827034, \hspace{0.5cm} P_{({\rm iii})}=1,\nonumber\\
U_{({\rm iv})}&=&0.00453444, \hspace{0.5cm} P_{({\rm iv})}\simeq 0.00899826.
\end{eqnarray}
Note that the uncertainty product and the probability for the case (iv) in (4.4) and (4.5) are approximate ones.
\\

The uncertainty product $U$ satisfies the Kennard relation $U\geq \frac{1}{2}$ for the cases (i) and (iii), for which the quantum mechanical 
probability $P=1$. On the other hand, the uncertainty product $U$ is clearly smaller than the lower bound of the Kennard relation $U\ll \frac{1}{2}$ for the specific samplings of the data in (ii) and (iv), for which the quantum mechanical probability is also very small $P\ll 1$.
  
The agreement of the uncertainty product $U_{({\rm iii})}$ for the above two cases with $N=100$ and $N=200$ indicates that the state after the measurement, namely, the 
step function-type wave function is universal to a good accuracy, as it should be. In our simple examples, the ratio $P/U$ is always of the order of unity. See also (2.4) and (2.5).
\\

\noindent {\bf Details of the numerical calculation for $n=10$ and $N=200$:}\\

We now explain the details of the numerical calculation for the specific 
case with the parameters $n=10$ and $N=200$.
 
In Fig.1, we have shown the distribution $|\psi_{n,1}(\bar{x},0)|^{2}$ for $n=10$. For this wave function we have
\begin{eqnarray}
&&(\Delta \bar{x})_{({\rm i})}=0.180756,\ (\Delta \bar{p})_{({\rm i})}=3.14159,\nonumber\\
&&U_{({\rm i})}=(\Delta \bar{x})_{({\rm i})}(\Delta \bar{p})_{({\rm i})}=0.567862
\end{eqnarray}
and $P_{({\rm i})}=1$ since we accept all the measured results without any bias.

In Fig.2, we show $|\phi_{l_{0}}(\bar{x},0)|^{2}$ for $l_{0}=80$. 

\begin{figure}[ht]
  \begin{center}
    \includegraphics[width=7cm]{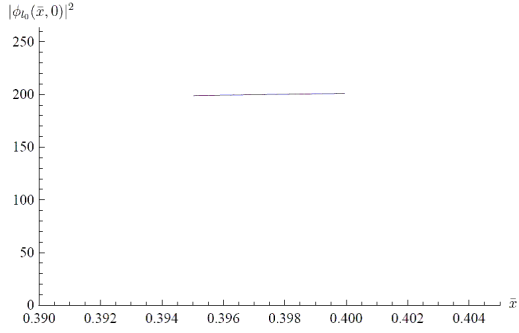}
  \caption{The distribution $|\phi_{l_{0}}(\bar{x},0)|^{2}$ with $l_{0}=80$ for $n=10$}
  \end{center}
\end{figure}
In Fig. 3, we show $|a_{l_{0},k}|^{2}$ for $l_{0}=80$, which shows that the cut-off at 
$k=4N=800$ is reasonable. 
\begin{figure}[ht]
  \begin{center}
    \includegraphics[width=7cm]{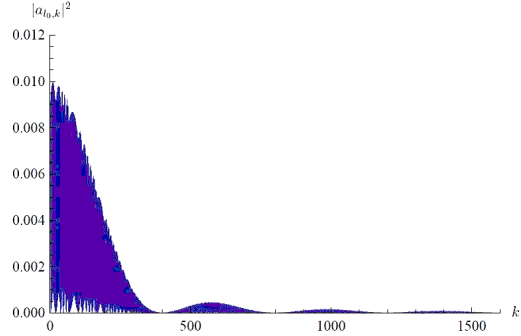}
  \caption{$|a_{l_{0},k}|^{2}$ for $l_{0}=80$}
  \end{center}
\end{figure}
To check this cut-off we show 
\begin{eqnarray}
|\sum_{k=1}^{800}a_{l_{0},k}\psi_{n,k}(\bar{x},0)|^{2} 
\end{eqnarray}
in Fig.4, which is to be compared to $|\phi_{l_{0}}(\bar{x},0)|^{2}$. These two should agree to a good accuracy if our approximation is valid.

\begin{figure}[ht]
  \begin{center}
    \includegraphics[width=7cm]{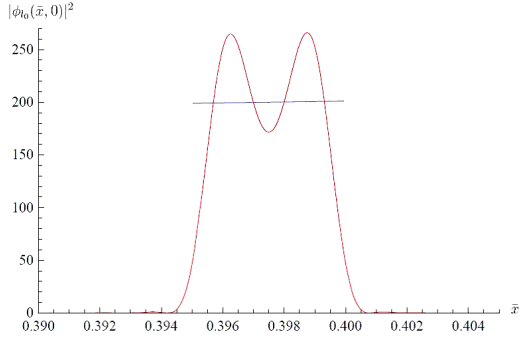}
  \caption{The series (4.7), which is cut-off at $k=4N=800$, is compared to $|\phi_{l_{0}}(\bar{x},0)|^{2}$ in Fig. 2.}
  \end{center}
\end{figure}

From Fig. 2, we have
\begin{eqnarray}
(\Delta \bar{x}_{l_{0}})_{({\rm ii})}=0.00144336.
\end{eqnarray}
In comparison we have $(\Delta \bar{x}_{l_{0}})_{({\rm ii})}=0.0013598$, which is close to the value in (4.8), from the series cut off at $k=800$ 
\begin{eqnarray}
\sum_{k=1}^{800}a_{l_{0},k}\psi_{n,k}(\bar{x},0) 
\end{eqnarray}
shown in Fig.4. To be precise, we use the normalized function in our numerical evaluation
\begin{eqnarray}
\frac{\sum_{k=1}^{800}a_{l_{0},k}\psi_{n,k}(\bar{x},0)}{ \sqrt{\int_{0}^{1}|\left(\sum_{k=1}^{800}a_{l_{0},k}\psi_{n,k}(\bar{x},0)\right)|^{2}d\bar{x}}}=\frac{\sum_{k=1}^{800}a_{l_{0},k}\psi_{n,k}(\bar{x},0)}{ \sqrt{\sum_{k=1}^{800}|a_{l_{0},k}|^{2}}}
\end{eqnarray}
and also in Fig.4. 

We thus have the uncertainty product and the probability for the analysis (ii)
\begin{eqnarray}
&&U_{({\rm ii})}=(\Delta \bar{x}_{l_{0}})_{({\rm ii})}(\Delta \bar{p})_{({\rm i})}=0.00453444, \nonumber\\
&&P_{({\rm ii})}
= B_{l_{0}}/N=0.00899826
\end{eqnarray}
by using 
\begin{eqnarray}
B_{l_{0}}=\int_{0}^{1}d\bar{x}|\psi_{n,1}(\bar{x},0)u_{l_{0}}|^{2}=1.79965
\end{eqnarray}
for $l_{0}=80$ and $n=10$.

From Fig.3, the cut-off at $k=800$ is reasonable. We then have  
\begin{eqnarray}
(\Delta \bar{p}_{l_{0}})_{({\rm iii})}=572.993 
\end{eqnarray}
from the formulas in (4.1) and (4.2) with the cut-off at $k=800$.
We then have the uncertainty product for the analysis (iii)
\begin{eqnarray}
U_{({\rm iii})}=(\Delta \bar{x}_{l_{0}})_{({\rm ii})}(\Delta \bar{p}_{l_{0}})_{({\rm iii})}=0.827034
\end{eqnarray}
and $P_{({\rm iii})}=1$ since we accept all the events without any bias. 

Finally, we have the uncertainty product and the probability for the analysis (iv)
\begin{eqnarray}
U_{({\rm iv})}&=&(\Delta \bar{x}_{l_{0}})_{({\rm ii})}(\Delta \bar{p})_{({\rm i})}=0.00453444,
\nonumber\\ 
P_{({\rm iv})}&=&|a_{l_{0},1}|^{2}=P_{({\rm ii})}.
\end{eqnarray}
The value of $U_{({\rm iv})}$ is the same as that of $U_{({\rm ii})}$
by definition.

By this way, we reproduce the numerical results in (4.5).

\section{Discussion and Conclusion}

We have studied an interplay of  uncertainty and probability in quantum 
mechanics. If one samples a suitable set of measured data, one can realize a very small uncertainty product but the probability of such a sampling of preferred events is very small. This mechanism provides a consistent explanation of the evasion of the uncertainty relation noted in~\cite{ballentine, ozawa2003,ozawa2004} in the framework of quantum mechanics; such a probability is simply very small. See Appendix. If one measures those events which are realized with almost certainty for a given state vector, then the standard Kennard relation is satisfied. We have also presented an example of cyclic measurements where the state vector is restored approximately to its original state while a product of measured standard deviations of coordinate and momentum is much smaller than the lower bound of the Kennard relation. Again, the probability of such a sampling of events is very small.
 
The present analysis shows that it is indispensable to examine the quantum mechanical probability when a possible very small uncertainty product is discussed. The consistency of our analysis with the Landau-Pollak type uncertainty relation~\cite{miyadera2007}, which also emphasizes the probability aspect, was also noted.

We now briefly comment on  a possible practical implication of our analysis. It is known that the detection of gravitational waves involves 
very weak signals, and thus the precise analysis of the detection limit provided 
by quantum mechanics (and possibly the evasion of the ``standard quantum limit'')
is important \cite{maddox}. Some of the authors argued that such an evasion of the standard quantum limit is impossible \cite{braginsky, caves, caves1}, while  others argued that the evasion of the standard quantum limit is possible \cite{yuen, ozawa1988}. Our analysis operates entirely 
within the framework of quantum mechanics and thus no notion such as ``the evasion of the standard quantum limit'' appears. Nevertheless, our analysis shows that one needs to examine the quantum mechanical probability to observe
the gravitational waves in a specific  setting of the detector when one analyzes the evasion or observance of the standard quantum limit. A further refinement of our analysis which emphasizes the role of quantum probability in the analysis of the uncertainty relation may help reconcile the contradicting views about the standard quantum limit in the detection of gravitational waves\footnote{In the detection of gravitational waves, the precise measurement of the position and then how to control the subsequent time development of the system is essential.
In this respect, our analysis (ii) of the position measurement and  the (immediately) subsequent specification of the momentum distribution in the analysis (iv) in Section 3  may be relevant. In fact, Caves, who 
defends the existence of the standard quantum limit against the criticism
by Yuen\cite{yuen}, comments that ``The measurements suggested by Yuen are among those for which no realization is known'' in \cite{caves1}. This comment might have some connection with the present analysis of a possible construction of a very small uncertainty product but with a very small probability of realizing such a product.}.
\\

We thank S. Tanimura for helpful comments. We are grateful to an anonymous referee for bringing the Landau-Pollak type uncertainty relation to our attention.

\appendix

\section{Implications on the past analyses of the uncertainty 
relation}

We here briefly mention the implications of the  analyses in Section 3 on the past analyses of the ``evasion of the uncertainty relation to an arbitrary degree with very small probability''\cite{fujikawa2008}.\\

\subsection{ Diffraction process}

In the context of the diffraction process of Ballentine\cite{ballentine},
the detector placed  on the screen corresponds to the very small detector in the analysis (ii) in Section 3. The momentum uncertainty in the diffraction process  is theoretically estimated at  \cite{ballentine}
\begin{eqnarray}
\delta p\sim p_{0}\frac{\delta q}{L}
\end{eqnarray}
for a given uncertainty $\delta q$ of the coordinate measured by a small detector with size $\delta q$ placed on the screen at the distance q from its center; $p_{0}$ is the momentum of the incoming particle and $L$
is the distance between the pin-hole and the screen in the diffraction process. A characteristic feature of (A.1) is that the momentum uncertainty does not increase for smaller $\delta q$ but rather decreases. In the mathematical limit $\delta q\rightarrow 0$ with fixed 
$L$, the above momentum uncertainty is eventually overtaken  by the intrinsic uncertainty in the incoming momentum
\begin{eqnarray}
\delta p\simeq p_{0}\frac{\delta q}{L}+ \delta p_{0}\frac{q}{L}
\simeq \delta p_{0}\frac{q}{L}
\end{eqnarray}
and the uncertainty product is given by 
\begin{eqnarray}
\delta p\delta q\simeq\delta p_{0}\frac{q}{L}\delta q 
\end{eqnarray}
by choosing $\delta q$ sufficiently small. Since the uncertainty of the momentum in the preparation process which ensures the presence of the particle in between the pin-hole and the screen with a unit probability is estimated at $\delta p_{0}\sim \frac{\hbar}{L}$~\cite{fujikawa2010} (note that we assume the Kennard relation $\delta p_{0}L\sim \hbar$ for events with a unit probability), the uncertainty product (A.3) is written as
\begin{eqnarray}
\delta p_{0}\frac{q}{L}\delta q \sim \hbar\frac{q}{L}\frac{\delta q}{L}\sim\hbar\frac{\delta q}{L}\ll \hbar.
\end{eqnarray}
One may understand that a transition from the ``classical'' domain
(A.1) without $\hbar$ to the quantum domain $\delta p\simeq\delta p_{0}\sim \hbar/L$ with $\hbar$ took place.

On the other hand, the quantum mechanical probability to find the diffracted particle in the interval $\delta q$ on the screen is 
\begin{eqnarray}
|\psi(q)|^{2}2\pi q \delta q \sim \frac{q}{L}\frac{\delta q}{L}\sim \frac{\delta q}{L}
\end{eqnarray}
where $\psi(q)$ is the two-dimensional wave function on the screen. We assumed an annulus-shaped detector for simplicity. We note that $|\psi(q)|^{2}\sim 1/L^{2}$ when one normalizes the wave function on the entire screen. To justify the estimate of (relative) probability in (A.5), one may imagine to send $N_{0}$ collimated particles (one particle at a time) through the pin-hole toward the screen. All the $N_{0}$ particles will eventually arrive at the screen, but only the tiny fraction $\sim N_{0}(q\delta q/L^{2})$ will arrive at the specific detector we consider; this fraction agrees with the probability (A.5). If one should cover the screen by many small detectors and if one should accept all the events detected by any of the small detectors, one would detect all $N_{0}$ particles but the standard deviation $\delta q$ of the measured coordinate would then be $\delta q\sim L$ to be consistent with the Kennard relation $\delta p_{0}\delta q\sim \hbar$.

If one identifies $\delta q/L\sim 1/N$, these relations (A.4) and (A.5) precisely correspond to 
the relations  (3.28) and (3.26) in the analysis (ii) in Section 3, respectively, including the form of the uncertainty product (A.4) in terms of the guessed uncertainty in the prepared momentum $\delta p_{0}$ and the uncertainty $\delta q$ in the measured coordinate.
(One can confirm that $\delta p_{0}\sim \hbar/\delta q$ if one wants to 
have a unit probability $|\psi(q)|^{2}2\pi q \delta q\sim 1$
by choosing $\psi(q)$ suitably, and in this case the uncertainty product becomes $\delta p_{0}\delta q\sim \hbar$.)
\\

\subsection{ Measurement-disturbance relation\\ $\epsilon(x)\eta (p)\ll \hbar$}

In the context of the gedanken experiment of Ozawa (see Section 
9 in \cite{ozawa2003}) of a two-particle system, which is specified by $(x_{1},p_{1})$ and $(x_{2},p_{2})$, the result of the analysis (i) in Section 3 may be used to confirm that the particle 1 existed in the state represented by $\psi_{n,1}(\bar{x},0)$ in (3.18) in the interval $0 \leq \bar{x}_{1}\leq 1$. One may next {\em assume} that the particle 1 of a two-particle system in  \cite{ozawa2003} in fact belonged to a specific state 
$\phi_{l_{0}}(\bar{x},0)$ in the expansion in (3.22)
\begin{eqnarray}
\psi_{n,1}(\bar{x},0)
=\sum_{l=1}^{N}c_{l}\phi_{l}(\bar{x},0)
\end{eqnarray}
{\em without} measurement, in the sense that the detector parameter $N$ is arbitrary. The precise measurement of the position $x_{2}$ of the particle 2  in \cite{ozawa2003}, for which the momentum of the particle 1 is not disturbed $\eta(p_{1})=0$ (or more realistically $\eta(p_{1})\sim\hbar/L$ if one puts a particle in a box with size $L$~\cite{fujikawa2010}), may be regarded as the specification of the position of the very small detector in the analysis (ii) in Section 3. The choice of the wave function of the particle 1, for 
which $\langle (x_{1}-x_{2})^{2}\rangle<\alpha$ with an arbitrary small 
$\alpha$ \cite{ozawa2003}, is then regarded to correspond to the precise overlap of a very narrow state $\phi_{l_{0}}(\bar{x},0)$ and a very small position detector in 
the analysis (ii) in Section 3. Note that $\langle (x_{1}-M)^{2}\rangle=\langle (x_{1}-x_{2})^{2}\rangle<\alpha$ with $M$ standing for the meter observable of the measuring apparatus in \cite{ozawa2003}, and thus our model represents the essence of the precise measurement of the position of the particle 1 in \cite{ozawa2003}. 
The {\em a priori} probability of the coincidence of  the state picked up by the small detector, whose position is specified by the precisely measured value of $x_{2}$, with the {\em assumed} narrow state $\phi_{l_{0}}(\bar{x},0)$ of the particle 1 is then $1/N$; this probability agrees with our quantum mechanical probability of finding the assumed $\phi_{l_{0}}(\bar{x},0)$ in (A.6) when the prepared wave function $\psi_{n,1}(\bar{x},0)$ of the particle 1 spreads over the domain  $0 \leq \bar{x}\leq 1$. In this case the condition $\eta(p_{1})=0$ is naturally preserved even when the position of the particle 1 is  specified by the detector with arbitrary accuracy $\epsilon(\bar{x}_{1})\sim 1/N$
 by choosing $N$ large \cite{ozawa2003}, although such a probability is very small $\sim 1/N$.


\begin{thebibliography}{99}
\bibitem{heisenberg}
W. Heisenberg, Z. Phys. {\bf 43} (1927) 172.
\bibitem{neumann}
J. von Neumann, {\em Mathematical Foundations of Quantum 
Mechanics} (Princeton University Press, Princeton, 1955).
\bibitem{appleby1998}
D.M. Appleby, Int. J. Theor. Phys. {\bf 37} (1998) 1491.
\bibitem{ozawa2004}
M. Ozawa, Phys. Lett. A{\bf 320} (2004) 367.
\bibitem{busch2007}
P. Busch, T. Heinonen and P. Lahti, Phys. Rep. {\bf 452} (2007) 155.
\bibitem{miyadera2008}
T. Miyadera and  H. Imai, Phys. Rev. A{\bf 78} (2008) 052119.
\bibitem{kennard}
E.H. Kennard, Z. Phys. {\bf 44} (1927) 326.
\bibitem{robertson}
H.P. Robertson, Phys. Rev. {\bf 34} (1929) 163.
\bibitem{ballentine}
L.E. Ballentine, Rev. Mod. Phys. {\bf 42} (1970) 358.
\bibitem{ozawa2003}
M. Ozawa, Phys. Lett. A{\bf 318} (2003) 21.
\bibitem{fujikawa2008}
K. Fujikawa and K. Umetsu, Prog. of Theor. Phys. {\bf 120} (2008) 797.
\bibitem{maddox}
J. Maddox, Nature (London) {\bf 331} (1988) 559.
\bibitem{miyadera2007}
T. Miyadera and  H. Imai, Phys. Rev. A{\bf 76} (2007) 062108, and references therein.
\bibitem{fujikawa2010}
K. Fujikawa, ``Comment on the uncertainty relation with periodic boundary 
conditons'', to be published in Prog. of Theor. Phys., arXiv:1009.5820[quant-ph].
\bibitem{braginsky}
V.B. Braginsky and Y.I. Vorontsov, Sov. Phys. Usp. {\bf 17} (1975) 644.
\bibitem{caves}
C.M. Caves, K.S. Thorne, R.W.P. Drever, V.D. Sandberg, and M. Zimmermann,
Rev. Mod. Phys. {\bf 52} (1980) 341. 
\bibitem{yuen}
H.P. Yuen,  Phys. Rev. Lett. {\bf 51} (1983) 719.
\bibitem{caves1}
C.M. Caves, Phys. Rev. Lett. {\bf 54} (1985) 2465. 
\bibitem{ozawa1988}
M. Ozawa, Phys. Rev. Lett. {\bf 60} (1988) 385.



\end{thebibliography}
\end{document}